\documentclass[reprint,amsmath,amssymb,aps,superscriptaddress]{revtex4-2}
\usepackage{graphicx}
\usepackage[colorlinks=true,allcolors=blue]{hyperref}
\usepackage{orcidlink}

\begin{document}

\title{Noiseless Linear Amplification and Loss-Tolerant Quantum Relay using Coherent State Superpositions}

\author{Joshua J. Guanzon\orcidlink{0000-0002-9990-6341}}
\email{joshua.guanzon@uq.net.au}
\affiliation{Centre for Quantum Computation and Communication Technology, School of Mathematics and Physics, The University of Queensland, St Lucia, Queensland 4072, Australia}

\author{Matthew S. Winnel\orcidlink{0000-0003-3457-4451}}
\affiliation{Centre for Quantum Computation and Communication Technology, School of Mathematics and Physics, The University of Queensland, St Lucia, Queensland 4072, Australia}

\author{Austin P. Lund\orcidlink{0000-0002-1983-3059}}
\affiliation{Dahlem Center for Complex Quantum Systems, Freie Universit\"at Berlin, 14195 Berlin, Germany}
\affiliation{Centre for Quantum Computation and Communication Technology, School of Mathematics and Physics, The University of Queensland, St Lucia, Queensland 4072, Australia}

\author{Timothy C. Ralph\orcidlink{0000-0003-0692-8427}}
\affiliation{Centre for Quantum Computation and Communication Technology, School of Mathematics and Physics, The University of Queensland, St Lucia, Queensland 4072, Australia}

\date{\today}

\begin{abstract}

Noiseless linear amplification (NLA) is useful for a wide variety of quantum protocols. Here we propose a fully scalable amplifier which, for asymptotically large sizes, can perform perfect fidelity NLA on any quantum state. Given finite resources however, it is designed to perform perfect fidelity NLA on coherent states and their arbitrary superpositions. Our scheme is a generalisation of the multi-photon quantum scissor teleamplifier, which we implement using a coherent state superposition resource state. Furthermore, we prove our NLA is also a loss-tolerant relay for multi-ary phase-shift keyed coherent states. Finally, we demonstrate that our NLA is also useful for continuous-variable entanglement distillation, even with realistic experimental imperfections. 

\end{abstract}

\maketitle

\section{Introduction}

Coherent states, which can be approximated by laser light, already exhibit favourable quantum properties such as minimum uncertainty. It follows naturally that the ability to control an arbitrary superposition of coherent states would be powerful for a wide range of quantum protocols, such as for quantum computing~\cite{jeong2002efficient,ralph2003quantum,lund2008fault,marek2010elementary,mirrahimi2014dynamically}, quantum metrology~\cite{munro2002weak,gilchrist2004schrodinger,joo2011quantum,zhang2013quantum,genoni2019non}, and quantum communication~\cite{sangouard2010quantum,brask2010hybrid,yin2014long,yin2019coherent}.

One type of control which is essential for many quantum protocols is \textit{amplification}, which increases the average number of photons. This could be done in a variety of different ways~\cite{zavatta2011high,fiuravsek2009engineering,marek2010coherent,mcmahon2014optimal,zhao2017characterization,zhang2018photon,hu2019entanglement,fiuravsek2022teleportation}, however an amplifier is more useful for quantum protocols if it satisfies two properties. First, the amplifier is \textit{linear}, where it preserves phase relations and thus arbitrary superposition of the input state. Second, the amplifier is \textit{noiseless}, where it doesn't add any additional noise in comparison to the input state. This is especially important for a superposition of coherent states, which are known to be susceptible to losing their quantum interference fringes~\cite{yurke1986generating,teh2020overcoming}. Therefore, \textit{noiseless linear amplification} (NLA) produces the best quality amplified states, with the downside that in general this process must be probabilistic due to the no-cloning theorem~\cite{wootters1982single}.

A particular well-studied subset of coherent state superpositions is the cat state, which is an equal superposition of coherent states, and whose name is in reference to Schrodinger's cat thought experiment~\cite{Schrodinger_1935}. There are numerous known methods for generating high quality cat states~\cite{lund2004conditional,ourjoumtsev2006generating,glancy2008methods,gerrits2010generation,huang2015optical}. These cat states can then be used as a resource for NLA of an arbitrary superposition of coherent states, based on the procedure given in Ref.~\cite{neergaard2013quantum}. This process works by teleporting the input state onto a resource state, hence it is also known as teleamplification. This is an example of perfect fidelity NLA of a CV superposition state, using finite resources. More precisely, suppose we have a resource cat state with $N$-components, where $N$ is the number of coherent states in the superposition. We can use this teleamplifier to perform NLA on any arbitrary superposition state containing up to $N$-components. If applied to an input alphabet containing only coherent states, the teleamplifier still operates well in the presence of large losses, i.e., for use as an untrusted quantum repeater or relay; this could be useful for quantum key distribution (QKD) via $N$-ary phase-shift keyed ($N$-PSK) coherent states~\cite{hirano2003quantum,leverrier2009unconditional}. However, the results given in Ref.~\cite{neergaard2013quantum} only proves and provides the optical networks required to perform NLA on states with $N=2$ or $N=4$ components. 

In this work, we propose a fully scalable teleamplifier protocol, which can perform perfect fidelity NLA for any integer $N \geq 2$, thereby filling in this missing knowledge gap. As an immediate corollary, since any quantum state can be represented in the asymptotic limit of $N\to\infty$ components~\cite{janszky1993coherent}, our protocol can in principle perform NLA on any arbitrary quantum state. Indeed, our proposed device can actually implement the so called exact immaculate amplification process, first hypothesised in Ref.~\cite{pandey2013quantum}. The structure of our proposed device is inspired by the multi-photon quantum scissor teleamplifier. The single-photon quantum scissor~\cite{pegg1998optical} can perform perfect fidelity NLA on quantum states containing up to a single photon~\cite{ralph2009nondeterministic}. There were attempts to generalise it to multi-photons, but the output states were distorted~\cite{ralph2009nondeterministic,xiang2010heralded}. It was only recently that better methods were found, which allowed perfect fidelity NLA on states containing up to any chosen number of photons~\cite{winnel2020generalized,guanzon2022ideal,fiuravsek2022optimal,zhong2022quantum}. We will show that our proposed scheme can be thought of as a type of generalisation to the multi-photon quantum scissor in Ref.~\cite{guanzon2022ideal}, in that it can perform teleamplification without photon truncation. 

We begin in Section~\ref{sec:proof}, where we prove that our protocol of size $N$, can in principle perform perfect fidelity NLA of an arbitrary state with $N$-components. We also show our protocol still works with high fidelity in situations where it is misaligned with the input. In Section~\ref{sec:loss}, we show how our device can also act as a loss-tolerant relay, given we send only coherent states. In Section~\ref{sec:distillation}, we explore another application for continuous-variable entanglement distillation, which we show is useful even with realistic experimental imperfections. Finally, we conclude in Section~\ref{sec:con}.

\section{$N$-Components Cat Teleamplifier} \label{sec:proof}

\begin{figure}[htbp]
    \begin{center}
        \includegraphics[width=\linewidth]{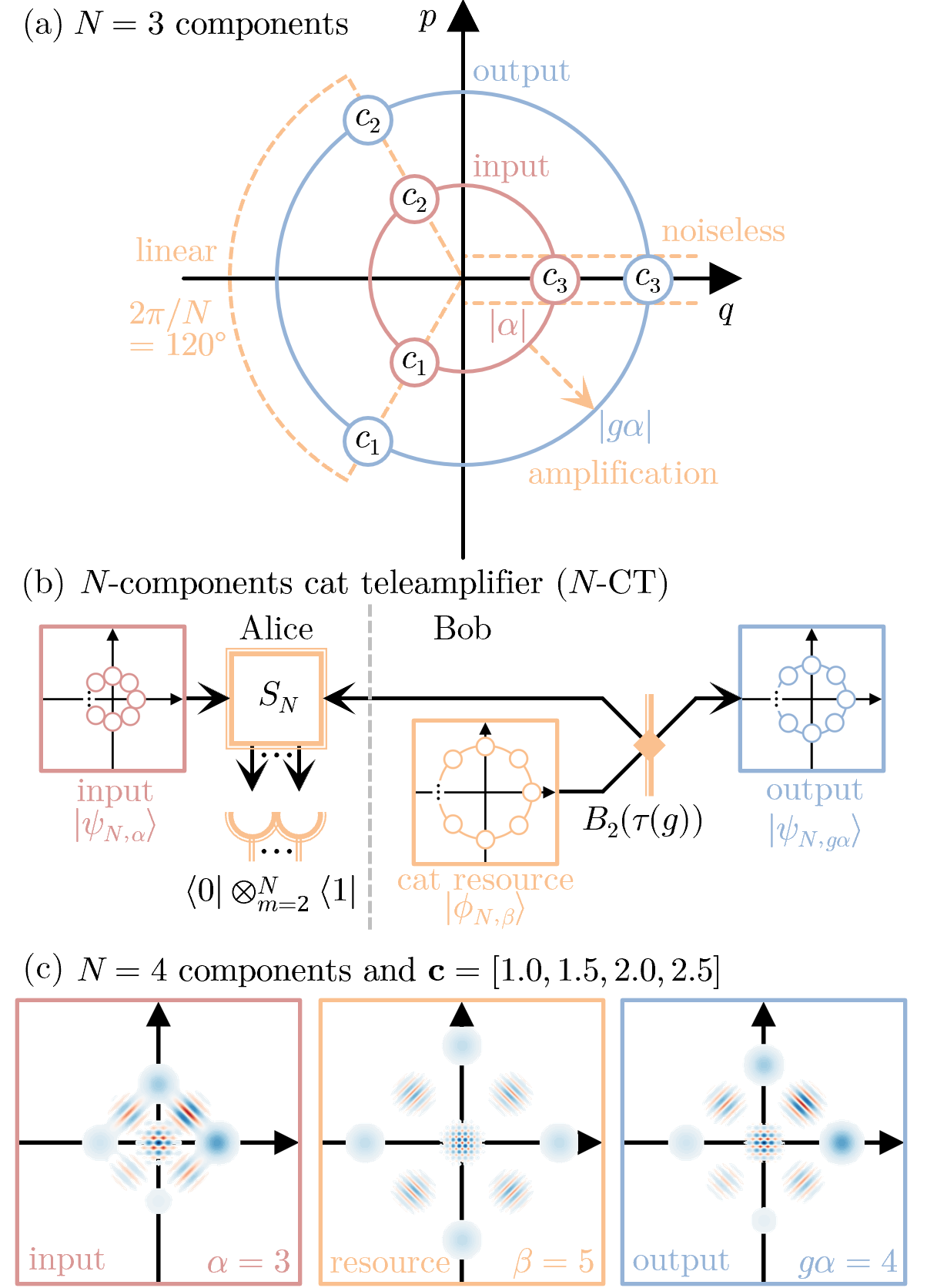}
        \caption{\label{fig:protocol} 
            Any arbitrary quantum state can be described by a continuous superposition of coherent states which lie on a phase space circle, since they form a complete basis~\cite{janszky1993coherent}. (a) We consider states made from finite $N$ coherent state components (here $N=3$), in a superposition with arbitrary $c_a$ coefficients $|\psi_{N,\alpha}\rangle \equiv \sum_{a=1}^{N} c_a |\omega_N^a\alpha\rangle$, with phase-shift angles $\omega_N\equiv e^{-i2\pi/N}$. An amplification (or de-amplification) process changes the circle radius from $|\alpha|$ to $|g\alpha|$ with $g\in[0,\infty)$ gain. (b) Our proposal performs noiseless linear amplification for any $N\in\mathbb{N}_{\geq2}$, with in principle perfect fidelity. Here Bob starts with an $N$-component cat state resource with amplitude $\beta=\alpha\sqrt{1+g^2}$, which is split on an unbalanced beam-splitter $B_2(\tau)$ with transmissivity $\tau=g^2/(1+g^2)$. Alice then mixes part of Bob's resource with the input state $|\psi_{N,\alpha}\rangle$ on a balanced $N$-splitter $S_N$. If Alice finds single-photons on all $N$ output ports except for the first port $\langle0|\otimes^N_{m=2}\langle1|$, then this heralds that Bob has an amplified version of the input state $|\psi_{N,g\alpha}\rangle$. (c) Phase space plots for $N=4$ coherent states given by blue blobs, with red striped interference fringes due to their superposition.}
    \end{center}
\end{figure}

Any quantum state can be represented as a superposition of Fock (photon number) states $|\psi\rangle = \sum_{n=0}^\infty c_n |n\rangle$, because these Fock states form a complete basis. Similarly, it is known that any quantum state can be represented as a continuous superposition of coherent states on a circle in phase space, because these coherent states also form a complete basis~\cite{janszky1993coherent}. In this regard, consider
\begin{align}
    |\psi_{N,\alpha}\rangle \equiv \sum_{a=1}^{N} c_a |\omega_N^a\alpha\rangle, \quad \omega_N\equiv e^{-i2\pi/N}, \label{eq:input}
\end{align}
which is an arbitrarily weighted $c_a$ superposition of $N$ coherent state components $|\omega_N^a\alpha\rangle$. These coherent states have the same magnitude $|\alpha|$, however the $a$th coherent state is rotated by an angle of $-2\pi a/N$. This is shown schematically for $N=3$ by the three smaller red circles in Fig.~\ref{fig:protocol}(a), which lie equally spaced on the perimeter of a larger red circle of radius $|\alpha|$. These states generalise the well-known cat state, therefore we will name this representation the cat basis. 

The NLA operation $g^{a^\dagger a}|\psi_{N,\alpha}\rangle \propto |\psi_{N,g\alpha}\rangle$, with gain $g\in[0,\infty)$, is shown by dashed yellow lines in Fig.~\ref{fig:protocol}(a). This transforms the red circle of radius $|\alpha|$ into the blue circle of radius $|g\alpha|$, while preserving all other properties. Our proposed protocol, with a scalable size parameter $N$, is given schematically in Fig.~\ref{fig:protocol}(b) in yellow. In this section, we will prove that this protocol can perform perfect fidelity NLA given the input state can be written as Eq.~\eqref{eq:input}. Hence, as an immediate corollary, our proposed protocol can in principle perform perfect fidelity NLA on any arbitrary input in the asymptotic limit of large $N$~\cite{janszky1993coherent}. Note we will show in a later section that even input states which can't be fully written as Eq.~\eqref{eq:input} can still experience good quality NLA using our protocol with finite $N$ sizes. 

Our protocol is powered by a resource of light called an $N$-components cat state
\begin{align}
    |\phi_{N,\beta}\rangle \equiv \frac{1}{\sqrt{\mathcal{N}}}\sum_{b=1}^{N} \omega^b_N |\omega_N^b\beta\rangle, 
\end{align}
where $\mathcal{N}$ is the normalisation constant. Cat states have been well studied, so there are many techniques to create these states~\cite{ourjoumtsev2007generation,takahashi2008generation,takeoka2008large,sychev2017enlargement}; the $N\in\{3,4\}$ sizes were first made experimentally a decade ago~\cite{vlastakis2013deterministically,kirchmair2013observation}. Due to this resource and the basis in which our scheme works, we will call our device the $N$-components cat teleamplifier ($N$-CT). 

Our device also requires standard linear optical components. We need a beam-splitter $B_2(\tau)$ with transmissivity $\tau\in[0,1]$, which scatters photons between two modes in a linear fashion as $a_1^\dagger \rightarrow \sqrt{\tau} a_1^\dagger - \sqrt{1-\tau}a_2^\dagger$. Note $a_m^\dagger$ is the creation operator which acts on the $m$th mode or port. We also need a balanced $N$-splitter $S_N$, which similarly has a linear action as $(a_1^\dagger,\ldots,a_N^\dagger)^T \rightarrow S_N (a_1^\dagger,\ldots,a_N^\dagger)^T,$ with a scattering matrix $(S_N)_{j,k} \equiv \omega^{(j-1)(k-1)}_N/\sqrt{N}$. Stated simply, a balanced $N$-splitter is just an $N$ modes generalisation of a balanced beam-splitter $S_{N=2} = B_2(\tau=1/2)$, with a particular phase configuration defined by the quantum Fourier transformation. Note that these linear transformations can be applied to coherent states by recalling the displacement operator definition $|\alpha\rangle \equiv e^{\alpha a^\dagger - \alpha^* a}|0\rangle$. 

We will now explain how our teleamplifier works in an ideal scenario. Bob uses the beam-splitter $B_2(\tau)$ to prepare the following state  
\begin{align}
    B_2(\tau) |0\rangle |\phi_{N,\beta}\rangle 
    &= \frac{1}{\sqrt{\mathcal{N}}} \sum_{b=1}^N \omega_N^b |-\omega_N^b\beta\sqrt{1-\tau}\rangle |\omega_N^b\beta\sqrt{\tau}\rangle \nonumber \\ 
    &= \frac{1}{\sqrt{\mathcal{N}}} \sum_{b=1}^N \omega_N^b |-\omega_N^b\alpha\rangle |\omega_N^bg\alpha\rangle. \label{eq:bobstate}
\end{align}
Bob purposefully chooses a particular amplification gain $g\in[0,\infty)$ by tuning the beam-splitter transmissivity to $\tau=g^2/(1+g^2)$, and preparing the cat resource state with an amplitude of $\beta=\alpha/\sqrt{1-\tau}=\alpha\sqrt{1+g^2}$. 

Bob then sends $|-\omega_N^b\alpha\rangle$ towards Alice, who mixes this state on the $N$-splitter $S_N$ with $|\psi_{N,\alpha}\rangle$ resulting in
\begin{align}
    &\sum_{a,b=1}^N \frac{c_a \omega_N^b}{\sqrt{\mathcal{N}}}  (S_N |\omega_N^a\alpha\rangle |-\omega_N^b\alpha\rangle \otimes^N_{m=3} |0\rangle) |\omega_N^bg\alpha\rangle \nonumber \\ 
    &= \sum_{a,b=1}^N \frac{c_a \omega_N^b}{\sqrt{\mathcal{N}}}  \otimes^N_{m=1} |(\omega_N^a-\omega_N^{b+m-1})\alpha/\sqrt{N}\rangle |\omega_N^bg\alpha\rangle. \label{eq:abstate}
\end{align}
Alice will then perform single-photon measurements on the $m\in\{1,\ldots,N\}$ output ports of the $N$-splitter. Notice that the $m_0=a-b+1$ output port is guaranteed to have no light since $|(\omega_N^a-\omega_N^{b+m_0-1})\alpha/\sqrt{N}\rangle=|0\rangle$. Put simply, the $b=a$ terms must measure no photons in the first mode $m_0=1$, while the $b\neq a$ terms must measure no photons in a different mode $m_0\neq 1$. Alice exploits this fact by selecting on the measurement outcome $\langle0|\otimes^N_{m=2}\langle1|$, in which only the $b=a$ terms have non-zero overlap, as follows
\begin{align}
    &\langle0|\otimes^N_{m=2}\langle1|\ \otimes^N_{m=1} |(\omega_N^a-\omega_N^{b+m-1})\alpha/\sqrt{N}\rangle \nonumber \\ 
    &= \delta_{b,a} \omega_N^{a(N-1)} \frac{e^{-|\alpha|^2} \alpha^{N-1}}{N^{(N-3)/2}}, 
\end{align}
where the magnitude is proven in Appendix~\ref{sec:ampproof}. 

The unnormalised output state after Alice measures $\langle0|\otimes^N_{m=2}\langle1|$ will then be 
\begin{align}
    |\psi_{N,g\alpha}\rangle = \frac{e^{-|\alpha|^2} \alpha^{N-1}}{N^{(N-3)/2}\sqrt{\mathcal{N}}} \sum_{a=1}^N c_a |\omega_N^a g\alpha\rangle. \label{eq:output}
\end{align}
Therefore, we have proven that our $N$-CT performs the required NLA operator $g^{a^\dagger a}$ with perfect fidelity, on a superposition of coherent states. In Fig.~\ref{fig:protocol}(c), we have plotted an example of the input, resource, and output superposition states in phase space. We have set the input state coefficients $\mathbf{c}\equiv[c_1,\ldots,c_N]$ such that each component clearly has a different weighting. We can see that our technique uses an equally weighted cat resource, to teleport the properties of the input state to the output state with a chosen amplification gain $g$.

This operation occurs with a success probability of 
\begin{align}
    \mathbb{P} = \langle \psi_{N,g\alpha}|\psi_{N,g\alpha} \rangle. \label{eq:prob}
\end{align}
This success probability can be improved by a factor of $N$, if Alice can select on any measurement of the form $\otimes_{m=1}^{m_0-1}\langle1| \otimes \langle0| \otimes_{m=m_0+1}^N\langle1|$, where $m_0\in\{1,\ldots,N\}$ is the mode that measured vacuum. Due to the symmetry of the balanced $N$-splitter, these measurements produce the same output states, but rotated by an angle of $2\pi(m_0-1)/N$. Therefore, this can be physically corrected via a feed-forward mechanism which applies a $\omega_N^{(m_0-1)a^\dagger a}$ phase shift. Alternatively, this can be virtually corrected if Bob is just going to measure the output state, by apply the required rotation on the results via software. A detailed analysis about these $N$ measurements which Alice can accept is in Appendix~\ref{sec:measureproof}. 

\begin{figure}[htbp]
    \begin{center}
        \includegraphics[width=\linewidth]{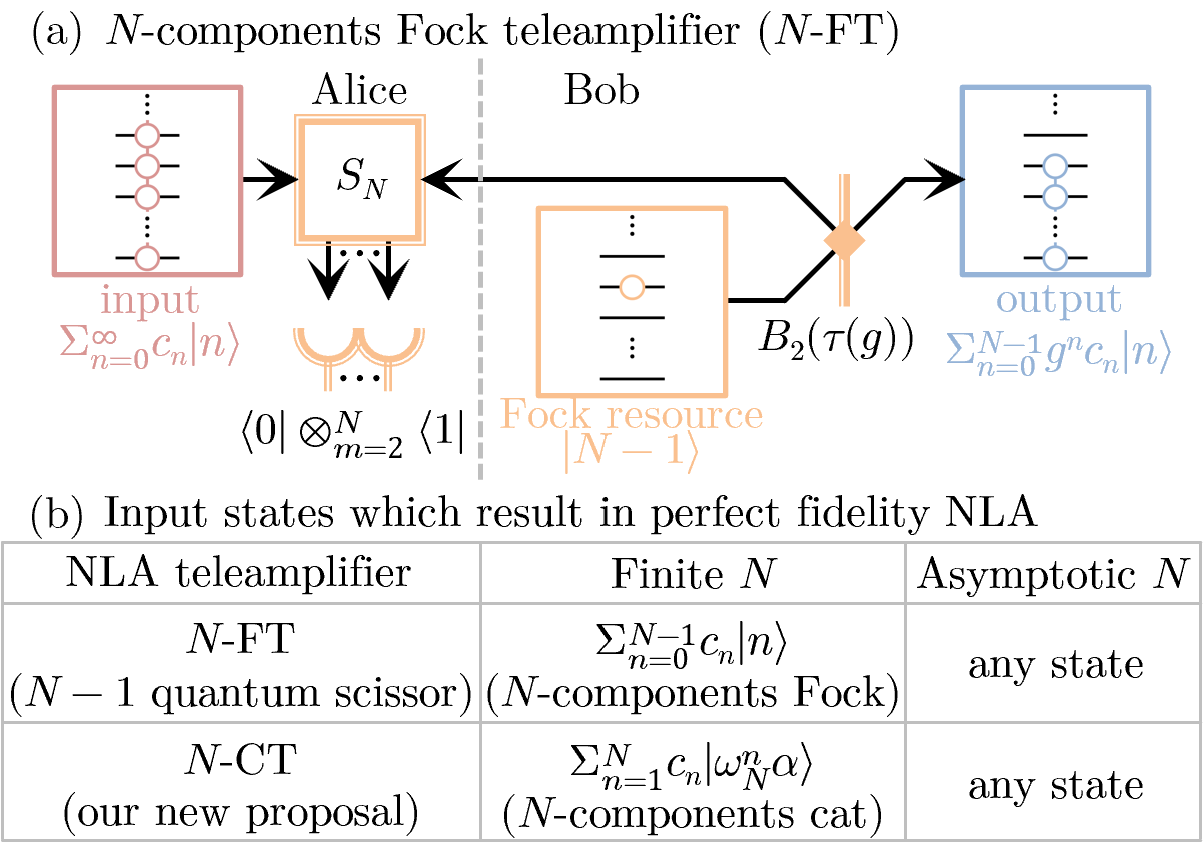}
        \caption{\label{fig:protocol-fock} 
            (a) Our proposed $N$-CT protocol is a generalisation of the $N$-components Fock teleamplifier, or more commonly known as the $(N-1)$-photons quantum scissor~\cite{guanzon2022ideal}. The only difference is that an $|N-1\rangle$ bunched photon state is used as the resource, instead of a $|\phi_{N,\beta}\rangle$ cat state; in fact, in the small amplitude $\beta\to0$ limit, these two resource states are the same. (b) These two teleamplifiers can NLA $N$-components in their particular basis with perfect fidelity. For asymptotically large $N$ these are complete bases, therefore in principle perfect fidelity NLA can be done for any quantum state.}
    \end{center}
\end{figure}

The structure of our $N$-CT device is similar to the $(N-1)$-photons quantum scissor protocol in Ref.~\cite{guanzon2022ideal}, shown schematically in Fig.~\ref{fig:protocol-fock}. This protocol is an $N$-components Fock teleamplifier ($N$-FT)~\cite{guanzon2022ideal}, in that it can perform NLA with perfect fidelity on any quantum state which can be written in the form $\sum_{n=0}^{N-1}c_n|n\rangle$. The only difference is that the $N$-FT is powered by a Fock state $|N-1\rangle$ as a resource of light. In fact, Ref.~\cite{janszky1995quantum} and Appendix~\ref{sec:catres} shows that for low $\beta$ amplitudes the cat state resource used in $N$-CT becomes the Fock state resource used in $N$-FT, where $\lim_{\beta\to0} |\phi_{N,\beta}\rangle = |N-1\rangle$. In this way, one may consider our $N$-CT as a type of generalisation of the $N$-FT in Ref.~\cite{guanzon2022ideal}. This resource generalisation allows our $N$-CT device the ability to amplify states without any Fock state truncation; a useful property to have if we want to preserve high photon correlations. If we assume a fixed amount of resource light $\beta$, the success probability of our $N$-CT scales with respect to gain as $\mathbb{P}\propto |\alpha|^{2(N-1)} \propto (1+g^2)^{-(N-1)}$. This is the same as the $N$-FT~\cite{guanzon2022ideal} and asymptotically equivalent to the theoretical maximum scaling~\cite{pandey2013quantum}, therefore we pay no significant success probability price for this generalisation. In fact, if we can modify the amount of resource light $\beta$, our $N$-CT can still have significant success probability in the limit of large gain $g$ (a feat which is not possible for the $N$-FT), which we will show later in the next section. 

\begin{figure}[htbp]
    \begin{center}
        \includegraphics[width=\linewidth]{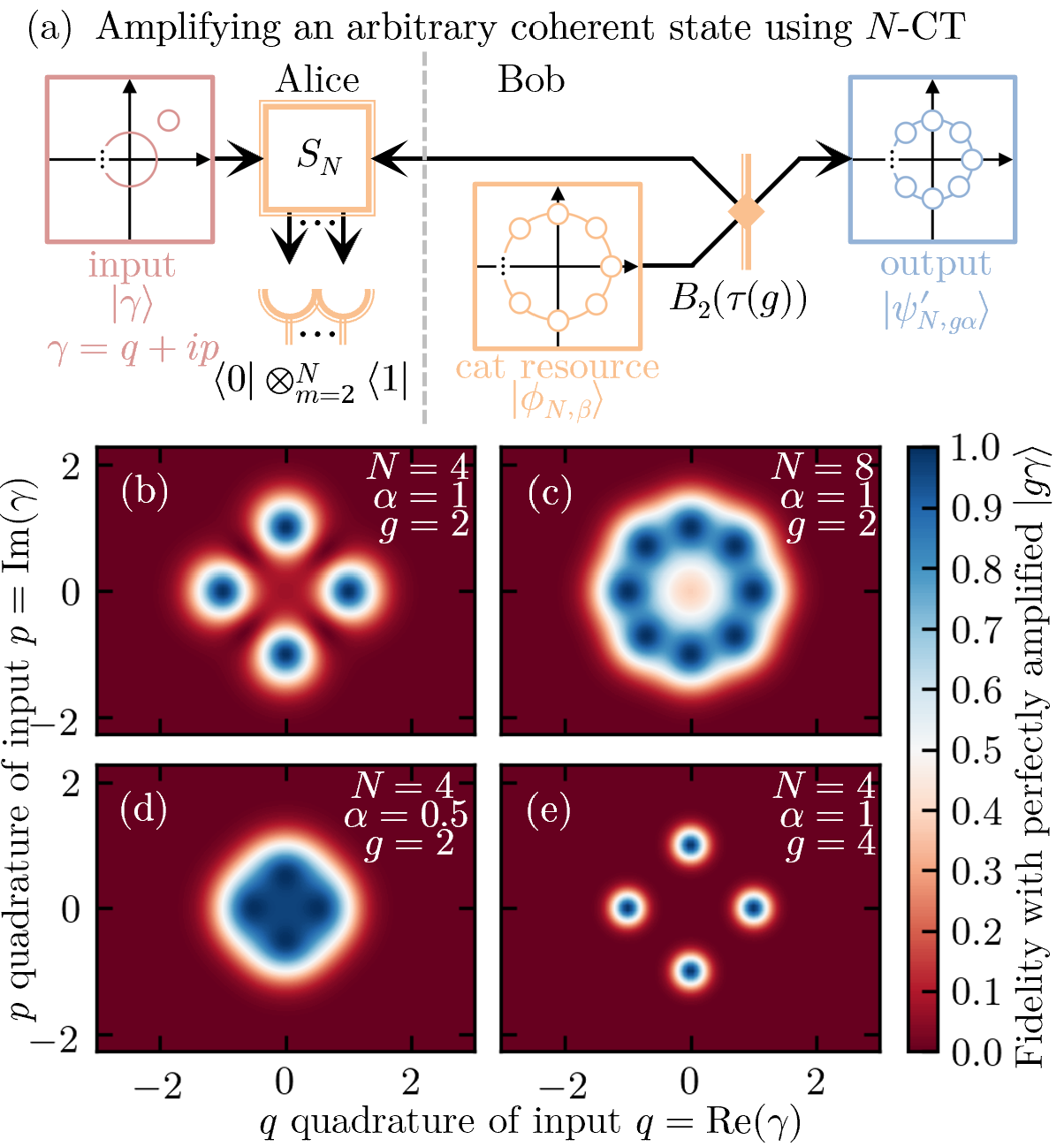}
        \caption{\label{fig:protocol-coherent} 
            Suppose we amplify an arbitrary coherent state input $|\gamma\rangle$ as shown in (a), and consider the fidelity against the related perfectly amplified state $|g\gamma\rangle$. We can see in (b) that it doesn't have to be exactly at the ideal place to be amplified with high fidelity. If we increase the protocol's size parameter $N$ as in (c), then it becomes more phase insensitive. This also happens if we reduce the protocol's expected amplitude parameter $\alpha$ as in (d), but at the cost of being less able to amplify larger states. Finally, if we demand more amplification gain $g$ as in (e), we require more knowledge about the input state to perform this with high fidelity.}
    \end{center}
\end{figure}

Let us consider what would happen if the input state $|\gamma\rangle$ is not exactly one of the chosen few coherent states (or any superposition of them), as shown in Fig.~\ref{fig:protocol-coherent}(a). In Appendix~\ref{sec:arbcoh} we prove that the output is still just an $N$-components coherent state superposition $|\psi'_{N,g\alpha}\rangle$, whose coefficients depend on the input parameter $\gamma$. We also derive an analytical expression for fidelity $F_\gamma \propto |\langle g\gamma|\psi'_{N,g\alpha}\rangle|^2$, which is a comparison with the output state from a perfect NLA $|g\gamma\rangle\propto g^{a^\dagger a}|\gamma\rangle$. This fidelity $F_\gamma$ is plotted in Fig.~\ref{fig:protocol-coherent}(b) to (e), which shows that our device can still amplify with high fidelity even if the input state is misaligned. Our protocol has the physical parameter set $\{N,\beta,\tau\}$, but for ease of understanding we can change this into a more pedagogical parameter set $\{N,\alpha,g\}$. We can interpret $|\alpha|$ as the protocol's expected input magnitude which is the large red circle in (a), while $|g\alpha|$ is the protocol's expected output magnitude which is the large blue circle. The actual input magnitude $|\gamma|$ amplifies well if it is near what is expected $|\alpha|$, as shown in (b) and (d). Interestingly, if we increase the gain $g$ this decreases the size of acceptable values, as shown when comparing (b) and (e). This makes sense, as any misalignment between the input and protocol will become more apparent at higher gain. In general, if we increase how crowded the coherent states are on the phase space circle, we increase the phase space insensitivity of our protocol.

\section{Loss-Tolerant Relay of Coherent States} \label{sec:loss}

\begin{figure}[htbp]
    \begin{center}
        \includegraphics[width=\linewidth]{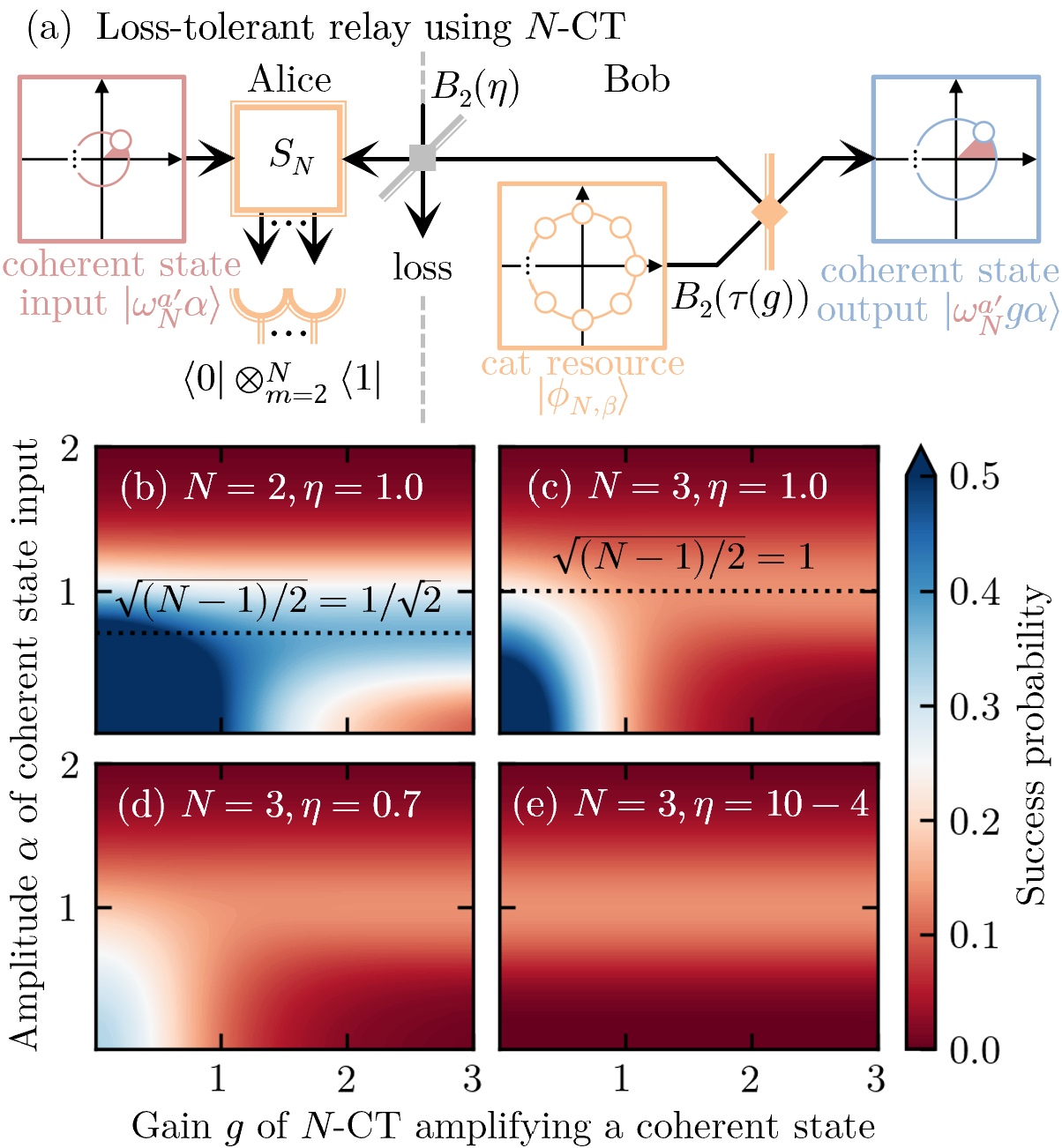}
        \caption{\label{fig:protocol-loss} 
            (a) Alice may choose to send only one coherent state $|\omega^{a'}_N\alpha\rangle$, out of the $N$ possible phase angles $a'\in\{1,\ldots,N\}$. This can be done despite limited transmissivity $\eta\in[0,1]$ of the channel connecting Alice and Bob (i.e. loss tolerance). Bob needs to take into account of this loss by using a larger cat resource state $\beta=\alpha\sqrt{1/\eta+g^2}$ and setting the beam-splitter $B_2(\tau)$ transmissivity to $\tau=\eta g^2/(1+\eta g^2)$. As demonstrated in (b) and (c), our $N$-CT can have good success probability even with asymptotically large gain $g\to\infty$; an input state with $\alpha_\text{max}=\sqrt{(N-1)/2}$ amplitude maximises the success probability. As shown in (c), (d), and (e), loss has minimal effect on probability if we are considering amplification $g>1$; this holds true even with asymptotically large loss $\eta\to0$.}
    \end{center}
\end{figure}

Suppose that Alice and Bob are now connected by a lossy fibre channel, with a transmissivity of $\eta\in[0,1]$, as shown in Fig.~\ref{fig:protocol-loss}. Loss can be described as another beam-splitter $B_2(\eta)$ attached to an extra environment mode. This results in the following shared state
\begin{align}
    &B_2(\eta) B_2(\tau) |0\rangle  |\phi_{N,\beta}\rangle |0\rangle \nonumber \\ 
    &= \frac{1}{\sqrt{\mathcal{N}}}\sum_{b=1}^N \omega_N^b |-\omega_N^b\alpha \rangle |\omega_N^bg \alpha \rangle |\omega_N^b\epsilon\rangle, \label{eq:bobstate_loss}
\end{align}
where $|\omega_N^b\epsilon\rangle$ describes the light lost to the environment. In contrast to the lossless scenario in Eq.~\eqref{eq:bobstate}, Bob must take the amount of loss into account to achieve a particular gain $g=\sqrt{\tau}/\sqrt{(1-\tau)\eta}$, which means setting his beam-splitter transmissivity to $\tau=\eta g^2/(1+\eta g^2)$. The amplitude of the cat state resource must also be larger to compensate for the loss $\beta = \alpha/\sqrt{(1-\tau)\eta} = \alpha\sqrt{1+\eta g^2}/\sqrt{\eta}$. Note that the environment mode has an amplitude of $\epsilon = \beta\sqrt{(1-\tau)(1-\eta)} = \alpha \sqrt{1-\eta}/\sqrt{\eta}$, however it is more important to notice that this error state is correlated in phase $\omega_N^b$. 

Now, consider Alice mixing $|-\omega_N^b\alpha \rangle$ with her state $|\psi_{N,\alpha}\rangle$ on the balanced $N$-splitter $S_N$, and performing $N$ single-photon measurements $\otimes_{m=1}^{m_0-1}\langle1| \otimes \langle0| \otimes_{m=m_0+1}^N\langle1|$. Notice that since Eq.~\eqref{eq:bobstate_loss} is similar to Eq.~\eqref{eq:bobstate} but with an extra error mode, the output state is similar to Eq.~\eqref{eq:output} as follows
\begin{align}
    \frac{e^{-|\alpha|^2} \alpha^{N-1}}{N^{(N-3)/2}\sqrt{\mathcal{N}}} \sum_{a=1}^N c_a |\omega_N^a g\alpha\rangle |\omega_N^a\epsilon\rangle. \label{eq:output_loss}
\end{align}
This output state is unfortunately entangled with the environment mode. Therefore, if Alice chooses to send any entangled state of the form in Eq.~\eqref{eq:input}, the output Bob receives will have decoherence type errors (in which cat states are particularly vulnerable). However, if Alice instead chooses to send just one coherent state $|\omega_N^{a'} g\alpha\rangle$ (i.e. $c_a=\delta_{a,a'}$), then the output state will be 
\begin{align}
    |\omega_N^{a'} g\alpha\rangle |\omega_N^{a'}\epsilon\rangle,
\end{align}
which is separable from the environment (note this is also the case if Alice sends a mixture of coherent states). In other words, Alice can send information to Bob in a loss-tolerant manner, by encoding information via the phase-shift of coherent states $a'\in\{1,\ldots,N\}$. Note that with $N=2$, only binary information can be sent. Our fully scalable $N$-CT protocol means this can now be done with any number of phases $N$, which means $N$-ary information can now be sent in a loss-tolerant manner. 

This protocol is not only loss-tolerant in terms of output state fidelity, but also success probability. If we substitute in $c_a=\delta_{a,a'}$ into the unnormalised output state in Eq.~\eqref{eq:output_loss}, we can get the following success probability  
\begin{align}
\mathbb{P}_\text{c} &= \frac{e^{-2|\alpha|^2} |\alpha|^{2(N-1)}}{N^{(N-3)}\mathcal{N}}, \\ 
\mathcal{N} &= \sum_{j,k=1}^N \omega_N^{k-j} e^{(\omega_N^{k-j}-1)|\alpha|^2(1/\eta+g^2)}.
\end{align} 
We have plotted $N\mathbb{P}_\text{c}$ as contour graphs in Fig.~\ref{fig:protocol-loss} for particular values of $\{N,\alpha,g,\eta\}$, which shows that even large gain and/or large loss can still have good success probability. This holds true even asymptotically with $g\to\infty$ and $\eta\to0$, because physically the resource light is increased $\beta = \alpha\sqrt{1/\eta+g^2}$ to achieve the necessary gain and compensate for any loss (hence $\mathcal{N}\to N$ in these limits). In contrast, the $N$-FT uses a fixed resource state which means its success probability will always scale as $(1+g^2)^{-(N-1)}$~\cite{guanzon2022ideal}. Hence our proposed resource generalisation via the $N$-CT results in a significant success probability advantage. Note that the maximum success probability in these asymptotic limits is $\frac{2eN^2}{N-1} \left( \frac{N-1}{2eN} \right)^N$, which occurs with $\alpha_\text{max}=\sqrt{(N-1)/2}$ amplitude inputs. More detailed analysis on the success probability can be found in Appendix~\ref{sec:prob}.

Recall that this protocol increases the coherent state magnitude without changing the uncertainty profile, therefore this has applications towards discriminating between various overlapping nonorthogonal states~\cite{nair2012symmetric,becerra2013experimental}. Our teleamplifier could also be useful for QKD purposes via multi-arrayed phase shift keyed ($N$-PSK) coherent states. For example, the B92 protocol~\cite{bennett1992quantum} could be done using $2$-PSK coherent states~\cite{koashi2004unconditional,tamaki2009unconditional}, while the BB84 protocol~\cite{BENNETT20147} could be done using $4$-PSK coherent states~\cite{neergaard2013quantum,lo2007security}. One may consider the in-between $3$-PSK coherent states case, which is secure for CV QKD~\cite{bradler2018security}. Finally, it is known for arbitrary $N$-PSK coherent states, with CV QKD and reverse reconciliation, that increasing $N$ improves the secret key rate~\cite{sych2010coherent}. However, whether these security proofs and rate details hold with a post-selected amplifier is unknown, hence we will leave QKD applications as an open question for future research. 

Lastly, note that coherent states put through a pure loss channel causes a decrease in amplitude without any change in noise profile or phase angle. Hence it is possible to put another loss channel between the input and the $S_N$ component in Fig.~\ref{fig:protocol-loss}, without changing the output (besides a reduction in amplitude). This means if we want to transfer these coherent states over a particular length of optical fibre, then it is a good idea to put $S_N$ in the middle of the fibre, as this would mean the loss before and after $S_N$ is balanced. This set-up is extremely useful for quantum relay purposes. In fact, the $N=1$ CT can overcome the repeaterless bound~\cite{pirandola2017fundamental}, which sets the benchmark for quantum repeaters, without requiring quantum memories~\cite{lucamarini2018overcoming,winnel2021overcoming}.

\section{Continuous-Variable Entanglement Distillation} \label{sec:distillation}

\begin{figure}[htbp]
    \begin{center}
        \includegraphics[width=\linewidth]{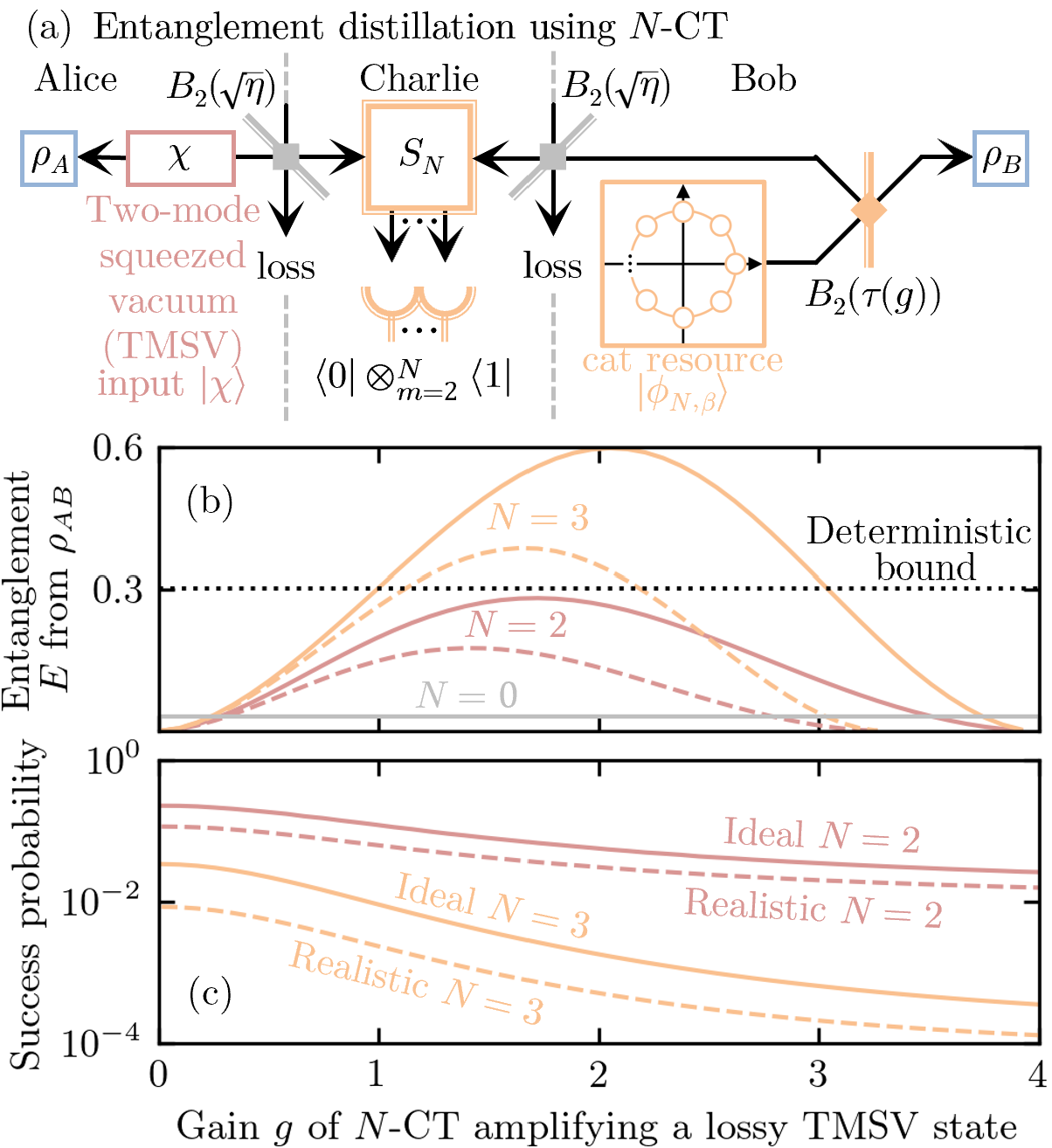}
        \caption{\label{fig:protocol-entanglement} 
            Suppose Alice wants to share a high quality entangled state with Bob, however they are connected by an imperfect channel which causes $95\%$ loss (i.e. $\eta=0.05$ transmissivity). If Alice sends one arm of a moderately squeezed TMSV state $|\chi=0.25\rangle$ to Bob, the resultant entanglement $E$ will be limited as shown by the solid gray $N=0$ line in (b). Note this entanglement is the Gaussian entanglement of formation $E$. Even if Alice starts with an infinitely squeezed TMSV state $|\chi=1\rangle$, the loss will still greatly limit the amount of entanglement, as shown by the dotted black line; this is called the deterministic bound. Our proposed $N$-CT in (a) can be used for probabilistic entanglement distillation. This recovers entanglement above the deterministic bound, depending on the size of our protocol $N$ as shown in (b). This holds even if we consider realistic experimental imperfections on the detectors and cat resource, as shown by the dashed lines. However, this additional entanglement comes at a cost in that the $N$-CT is probabilistic, with a success probability shown in (c).}
    \end{center}
\end{figure}

Quantum entanglement is a useful resource for many protocols. However, maintaining this entanglement from environmental loss and other imperfections is a major challenge. To make this more concrete, consider a scenario where Alice has a two-mode squeezed vacuum (TMSV) state $|\chi\rangle$. A particular measure of continuous-variable entanglement is the Gaussian entanglement of formation $E$~\cite{bennett1996mixed,wolf2004gaussian,marian2008entanglement}, which can be calculated numerically using the quantum state's covariance matrix~\cite{tserkis2017quantifying,tserkis2019quantifying}. If Alice's TMSV state is moderately squeezed by $\chi=0.25$, then initially the amount of entanglement between the two modes of this state is $E(\chi=0.25,\eta=1)\approx0.36$. However, suppose Alice sends one mode of the TMSV state through a channel to Bob, but unfortunately this channel only has a transmissivity of $\eta=0.05$ (meaning that any light put through it will experience $95\%$ loss). This resultant lossy TMSV state $\rho_{AB}$ now only has $E(\chi=0.25,\eta=0.05)\approx0.03$ entanglement, as indicated by the $N=0$ labelled solid gray line in Fig.~\ref{fig:protocol-entanglement}(b). Even if Alice starts off with an infinitely squeezed state $\chi=1$ with infinite entanglement, this loss limits the entanglement to only $E(\chi=1,\eta=0.05)\approx0.30$, as indicated by the horizontal dotted black line. This quantity is called the deterministic bound, as it is the best you can do without a probabilistic entanglement recovery process.

Our $N$-CT device can be used to probabilistically distill entanglement through a lossy channel, as shown schematically in Fig.~\ref{fig:protocol-entanglement}(a). We have positioned the $N$-splitter $S_N$ in the middle of the channel as this greatly improves the success probability scaling to $O(\sqrt{\eta})$. We choose a particular gain $g$ and set Bob's beam-splitter transmissivity to $\tau=g^2/(1+g^2)$. Note that this lossy TMSV input state can't be fully written in the finite $N$-components cat basis form given in Eq.~\eqref{eq:input}. Therefore, we simply optimise the cat resource amplitude $\beta$ for each gain $g$ to maximise the amount of entanglement $E$.

The resultant entanglement distillation for $\chi=0.25$, using $N=2$ and $N=3$ sizes is given by the solid red and yellow lines respectively in Fig.~\ref{fig:protocol-entanglement}(b). If we include experimental imperfections, modelled by $30$\% extra loss on the cat state resource and $70$\% efficiency single-photon detectors~\cite{reddy2019exceeding,lita2008counting,marsili2013detecting,miller2003demonstration}, we get the dashed lines. Even accounting for realistic experimental imperfections, we can see that increasing the size $N$ increases the entanglement. The reason for the bell shaped curves could be understood by considering how the transmissivity $\tau$ of the beam-splitter $B_2(\tau)$ changes with gain $g$. Small gain $g\to0$ means limited transmissivity $\tau\to0$, which results in limited amount of light exiting on Bob's side. On the other hand, large gain $g\to\infty$ means almost complete transmissivity $\tau\to1$, which results in not much of the resource light being entangled with Alice's input. From this physical intuition, it is clear that there should be an optimal gain $g$ in which the amount of entanglement $E$ is maximised. 

This extra entanglement comes at the cost of success probability as shown in Fig.~\ref{fig:protocol-entanglement}(c). However, our generalised $N$-CT device can beat the deterministic bound by using higher $N$ values, thus demonstrating it's usefulness for generating high-quality entanglement. This also clearly shows that our $N$-CT device is useful even if the input state doesn't fully satisfy Eq.~\eqref{eq:input}. We have provided all our simulation code and data in Ref.~\footnote{See \href{https://github.com/JGuanzon/cat-teleamplifier}{https://github.com/JGuanzon/cat-teleamplifier} for the $N$-CT simulation code and generated data which can recreate the graphs in this paper. This uses the Strawberry Fields python library, which includes Ref.~\cite{killoran2019strawberry,bromley2020applications,bourassa2021fast}}. 

\section{Conclusion} \label{sec:con}

We have proven that our $N$-CT protocol, given in Fig.~\ref{fig:protocol}(b), can be used to implement the noiseless linear amplification operator $g^{a^\dagger a}$. This can be done with perfect fidelity, if the input state can be fully written in the $N$-components cat basis as Eq.~\eqref{eq:input}. Since this basis is complete given asymptotic number of components $N\to\infty$, this means the $N$-CT can in principle perform perfect fidelity amplification on any quantum state. These results can be understood as an extension to the generalised quantum scissor $N$-FT protocol~\cite{guanzon2022ideal}, where the output state has no Fock truncation. We also demonstrated that our $N$-CT can still work with high fidelity even if the input doesn't align exactly with Eq.~\eqref{eq:input}. Furthermore, we have shown that if the input is one out of a set of $N$ coherent states, then amplification can be done in a loss-tolerant manner without decoherence, and with a significant success probability even in the large gain and large loss asymptotic limit. Therefore, our proposal also has uses for many quantum protocols which employee $N$-ary phase-shift keyed coherent states. Finally, we have also shown that our proposed $N$-CT device for finite $N$ is able to distil high quality continuous-variable entanglement through lossy channels, even assuming realistic experimental imperfections.

\begin{acknowledgments}
APL acknowledges support from BMBF (QPIC) and the Einstein Research Unit on Quantum Devices. This research was supported by the Australian Research Council Centre of Excellence for Quantum Computation and Communication Technology (Project No. CE170100012).
\end{acknowledgments}

\appendix

\section{Proof of Measurement Amplitude} \label{sec:ampproof}

The measurement amplitude resolves to 
\begin{align}
    &\langle0|\otimes^N_{m=2}\langle1|\ \otimes^N_{m=1} |(\omega_N^a-\omega_N^{b+m-1})\alpha/\sqrt{N}\rangle \nonumber \\ 
    &= \delta_{b,a} \langle0|\otimes^N_{m=2}\langle1|\ \otimes^N_{m=1} |(1-\omega_N^{m-1})\omega_N^a\alpha/\sqrt{N}\rangle \nonumber \\ 
    &= \delta_{b,a} e^{-\sum_{m=2}^N |1-\omega^{m-1}_N|^2|\alpha|^2/2N} \prod_{m=2}^N \frac{(1-\omega_N^{m-1})\omega_N^a\alpha}{\sqrt{N}} \nonumber \\ 
    &= \delta_{b,a} \omega_N^{a(N-1)} \frac{e^{-|\alpha|^2} \alpha^{N-1}}{N^{(N-1)/2}} \prod_{m=2}^N (1-\omega_N^{m-1})  \nonumber \\ 
    &= \delta_{b,a} \omega_N^{a(N-1)} \frac{e^{-|\alpha|^2} \alpha^{N-1}}{N^{(N-3)/2}}. \label{eq:app_ampproof}
\end{align}
We will justify all the critical steps of this derivation. The first equality in Eq.~\eqref{eq:app_ampproof} uses the fact that the state $\otimes^N_{m=1} |(\omega_N^a-\omega_N^{b+m-1})\alpha/\sqrt{N}\rangle$ has vacuum in the wrong mode other than $b=a$, as explained in detail in the main text after Eq.~\eqref{eq:abstate}. The second equality in Eq.~\eqref{eq:app_ampproof} used the representation of coherent states in the Fock basis $|\alpha\rangle = e^{-|\alpha|^2/2}\sum_{n=0}^\infty \alpha^n/\sqrt{n!} |n\rangle$. 

The third equality in Eq.~\eqref{eq:app_ampproof} uses the following to simplify the exponent 
\begin{align}
    \sum_{m=2}^N |1-\omega_N^{m-1}|^2 &= \sum_{m=1}^N |1-\omega_N^{m-1}|^2 \nonumber \\ 
    &= \sum_{m=1}^N (1-\omega_N^{m-1})(1-\omega_N^{-m+1}) \nonumber \\ 
    &= \sum_{m=1}^N (2-\omega_N^{m-1}-\omega_N^{-m+1}) \nonumber \\ 
    &= 2N, \label{eq:app_exponent}
\end{align}
where we used the geometric summation equation $\sum_{m=1}^N \omega_N^{m-1} = (1-\omega_N^N)/(1-\omega_N) = 0$, and similarly for $\sum_{m=1}^N \omega_N^{-m+1}$. 

The final equality in Eq.~\eqref{eq:app_ampproof} uses 
\begin{align}
    \prod_{m=2}^N (1-\omega_N^{m-1}) = \prod_{m=1}^{N-1} (1-\omega_N^m) = N, \label{eq:app_prod_0}
\end{align}
which we will derive based on the fact that $\omega_N\equiv e^{-2i\pi/N}$ are roots of unity. Consider the following polynomial
\begin{align}
    p(z) = z^N - 1. 
\end{align}
The condition $p(z)=0$ is true when $z=e^{-2i\pi k/N}=\omega_N^k$ for $k\in\{0,\ldots,N-1\}$; in other words, $\omega_N^k$ are the unique roots of unity. Therefore, this polynomial can also be written in factorisation form using these roots
\begin{align}
    p(z) = \prod_{k=0}^{N-1} (z-\omega_N^k) = (z-1) \prod_{k=1}^{N-1} (z-\omega_N^k). \label{eq:app_prod_1}
\end{align}
Using some basic algebraic manipulation, this same polynomial can also be written as
\begin{align}
    p(z) &= (z+z^2+\cdots+z^N) - (1+z+\cdots+z^{N-1}) \nonumber \\ 
    &= (z-1)(1+z+\cdots+z^{N-1}). \label{eq:app_prod_2}
\end{align}
Thus, comparing Eq.~\eqref{eq:app_prod_1} and Eq.~\eqref{eq:app_prod_2} we can see that
\begin{align}
    \prod_{k=1}^{N-1} (z-\omega_N^k) = 1+z+\cdots+z^{N-1}. \label{eq:app_prod_3}
\end{align}
By substituting in $z=1$, the right-hand side of this expression is equivalent to $N$ since there are $N$ terms. Thus we have proven our required relation.

\section{Alice Multiple Measurements Proof} \label{sec:measureproof}

\begin{figure}[htbp]
    \begin{center}
        \includegraphics[width=\linewidth]{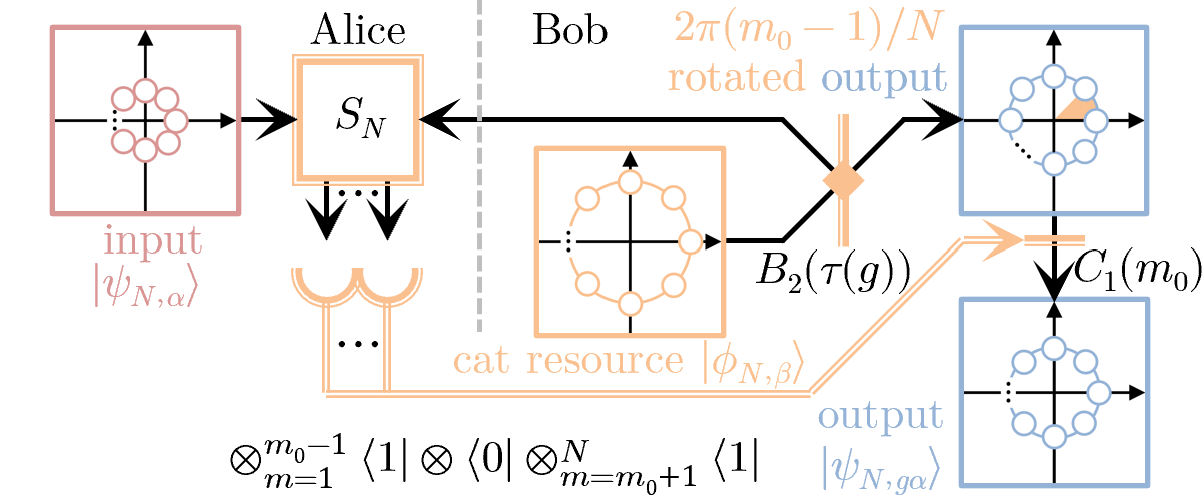}
        \caption{\label{fig:protocol-correction} 
            Alice is free to select on $N$ different measurement outcomes of the form $\otimes_{m=1}^{m_0-1}\langle1| \otimes \langle0| \otimes_{m=m_0+1}^N\langle1|$ for $m_0\in\{1,\ldots,N\}$. Note that $m=m_0$ is the output port or mode of the $N$-splitter $S_N$ that measured the vacuum outcome. Due to the symmetry of $S_N$, these measurements all produce the same output states but phase space rotated by $2\pi(m_0-1)/N$, which can be corrected as explained in text.}
    \end{center}
\end{figure}

We will prove here that Alice can select on a set of $N$ measurements $\otimes_{m=1}^{m_0-1}\langle1| \otimes \langle0| \otimes_{m=m_0+1}^N\langle1|$, and produce the same output state $|g\psi_{N,\alpha}\rangle$ but with a correctable phase-shift. Note that $m_0$ refers to the mode or output port of the balanced $N$-splitter which measured vacuum. 

Recall from Eq.~\eqref{eq:abstate} that the state Alice and Bob share just after the balanced $N$-splitter is
\begin{align}
    \sum_{a,b=1}^N c_a \omega_N^b \otimes^N_{m=1} |(\omega_N^a-\omega_N^{b+m-1})\alpha/\sqrt{N}\rangle |\omega_N^bg\alpha\rangle. \label{eq:app_abstate}
\end{align}
Each term $\otimes^N_{m=1} |(\omega_N^a-\omega_N^{b+m-1})\alpha/\sqrt{N}\rangle$ is guaranteed to measure vacuum in one output port $m=m_0$, which is when $\omega_N^a=\omega_N^{b+m_0-1}$ or $m_0=a-b+1$. 

Now, let us suppose Alice selects on a measurement where all output ports clicked with one photon, except for one port $m=m_0$. This could only have occurred due to the $b=a-m_0+1$ terms, because the $b\neq a-m_0+1$ terms are required to measure vacuum in mode $m\neq m_0$. In other words, all terms in the sum where $b\neq a-m_0+1$ have their vacuum state in the wrong mode to be able to satisfy the one photon detection measurements. Thus this measurement results in
\begin{align}
    &\otimes_{m=1}^{m_0-1}\langle1| \otimes \langle0| \otimes_{m=m_0+1}^N\langle1| \nonumber \\ 
    &\quad \otimes^N_{m=1} |(\omega_N^a-\omega_N^{b+m-1})\alpha/\sqrt{N}\rangle \nonumber \\
    &= \delta_{b,a-m_0+1} e^{-\sum_{m=1}^N|1-\omega_N^{m-m_0}|^2|\alpha|^2/(2N)} \nonumber \\ 
    &\quad \prod_{m=1}^{m_0-1} \frac{ (\omega_N^a-\omega_N^{a+m-m_0}) \alpha}{\sqrt{N}}  \prod_{m=m_0+1}^N \frac{ (\omega_N^a-\omega_N^{a+m-m_0}) \alpha}{\sqrt{N}} \nonumber \\
    &= \delta_{b,a-m_0+1} \omega_N^{a(N-1)} \frac{e^{-|\alpha|^2} \alpha^{N-1}}{N^{(N-1)/2}} \nonumber \\ 
    &\quad \prod_{m=1}^{m_0-1}(1-\omega_N^{m-m_0}) \prod_{m=m_0+1}^N(1-\omega_N^{m-m_0}) \nonumber \\ 
    &= \delta_{b,a-m_0+1} \omega_N^{a(N-1)} \frac{e^{-|\alpha|^2} \alpha^{N-1}}{N^{(N-3)/2}}. \label{eq:app_m0measure}
\end{align}
The second equality uses $\sum_{m=1}^N|1-\omega_N^{m-m_0}|^2=2N$ for the exponent, which can be proven like Eq.~\eqref{eq:app_exponent}. The last equality uses
\begin{align}
    &\prod_{m=1}^{m_0-1}\left(1 -\omega_N^{m-m_0} \right) \prod_{m=m_0+1}^N \left( 1-\omega_N^{m-m_0} \right) \nonumber \\ 
    &= \prod_{m=-m_0+1}^{-1}\left(1 -\omega_N^m \right) \prod_{m=1}^{N-m_0} \left( 1-\omega_N^m \right) \nonumber \\ 
    &= \prod_{m=N-m_0+1}^{N-1}\left(1 -\omega_N^m \right) \prod_{m=1}^{N-m_0} \left( 1-\omega_N^m \right) \nonumber \\ 
    &= \prod_{m=1}^{N-1} \left(1 -\omega_N^m \right) \nonumber \\ 
    &= N,
\end{align}
using $\omega_N^N=1$ and Eq.~\eqref{eq:app_prod_0}. 

Applying the result in Eq.~\eqref{eq:app_m0measure} to Eq.~\eqref{eq:app_abstate} produces the following output state
\begin{align}
    &\omega_N^{-m_0+1} \frac{e^{-|\alpha|^2} \alpha^{N-1}}{N^{(N-3)/2}\sqrt{\mathcal{N}}} \sum_{a=1}^N c_a  |\omega_N^{a-m_0+1}g\alpha\rangle \nonumber \\ 
    &= \omega_N^{(-m_0+1)(1+a^\dagger a)}|g\psi_{N,\alpha}\rangle. 
\end{align}
Therefore, we have shown irrespective of $m_0$, Bob can always recover the same output state by applying a phase shift correction of 
\begin{align}
    C_1(m_0)=\omega_N^{(m_0-1)a^\dagger a},
\end{align}
which we show schematically in Fig.~\ref{fig:protocol-correction}. Note this correction doesn't need to be implemented physically if Bob is simply measuring the output state, rather the required rotation correction can be implemented in software directly on Bob's measurement results. These measurements also have the same probability of success $\mathbb{P}=\langle g\psi_{N,\alpha} |g\psi_{N,\alpha}\rangle$, irrespective of $m_0$. Therefore, if all $N$ measurements can be accepted, then we can improve the success probability by a factor of $N$ to $N\mathbb{P}$.

\section{Cat Resource State in Fock Basis} \label{sec:catres}

We can represent the cat state resource in the Fock basis as follows
\begin{align}
    |\phi_{N,\beta}\rangle &\equiv \frac{1}{\sqrt{\mathcal{N}}} \sum_{b=1}^N \omega^b_N |\omega_N^b\beta\rangle \nonumber \\ 
    &= \frac{e^{-|\beta|^2/2}}{\sqrt{\mathcal{N}}} \sum_{b=1}^N \sum_{n=0}^\infty \omega^{b(n+1)}_N \frac{\beta^n}{\sqrt{n!}} |n\rangle,
\end{align}
where we used the coherent state representation $|\alpha\rangle = e^{-|\alpha|^2/2}\sum_{n=0}^\infty \alpha^n/\sqrt{n!} |n\rangle$. Note that most of these terms resolve to zero since 
\begin{align}
    \sum_{b=1}^N \omega^{b(n+1)}_N = \delta_{n+1,kN} N, \quad k\in\mathbb{N}. \label{eq:sumrootsunity}
\end{align}
This is because when $\omega^{b(n+1)}_N\neq1$ (i.e. $(n+1)\neq kN$), then the expression is the sum of all roots of unity which resolves to zero, or algebraically $\omega_N^{n+1}(1-\omega^{N(n+1)}_N)/(1-\omega^{n+1}_N)=0$. This means that our resource state can be simplified as 
\begin{align}
    |\phi_{N,\beta}\rangle &= \frac{Ne^{-|\beta|^2/2}}{\sqrt{\mathcal{N}}}  \sum_{k=1}^\infty  \frac{\beta^{kN-1}}{\sqrt{(kN-1)!}} |kN-1\rangle,
\end{align}
Notice that for very small amplitudes $\beta\ll 1$, the term with the largest coefficient will overwhelmingly be the $k=1$ term. From this, it is clear that for asymptotically small amplitudes this cat state reduces down to $\lim_{\beta\to0} |\phi_{N,\beta}\rangle = |N-1\rangle $. Note that this result was already known, as detailed in Ref.~\cite{janszky1995quantum}.

\section{Arbitrary Coherent State Input} \label{sec:arbcoh}

\begin{figure*}[htb]
    \begin{center}
        \includegraphics[width=\linewidth]{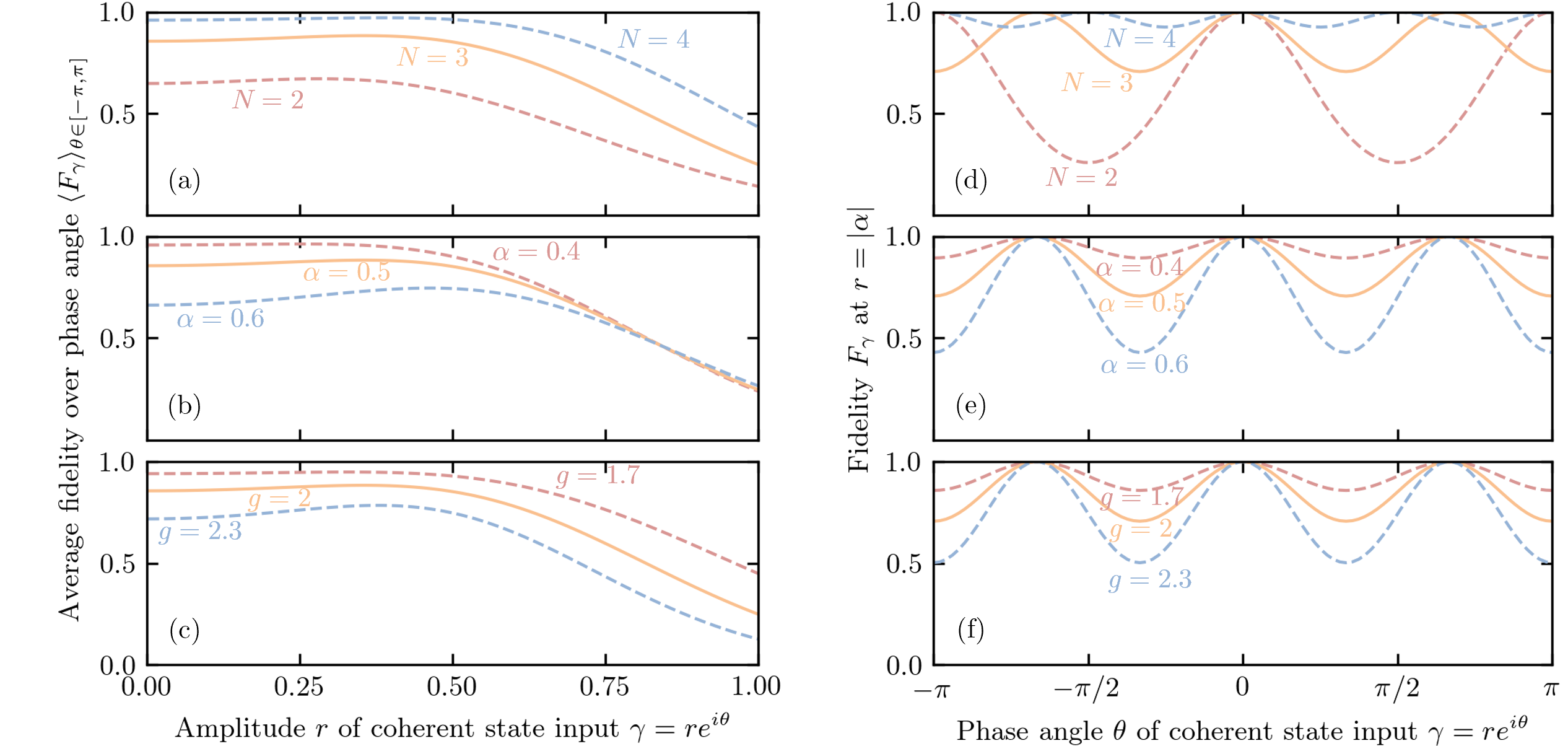}
        \caption{\label{fig:coherent-fid} 
            Here we use fidelity $F_\gamma$ as a measure for how well our device can amplify a completely arbitrary coherent state $|\gamma\rangle$ to $|g\gamma\rangle$. The left graphs, (a) to (c), is the fidelity for amplifying a coherent state with $r=|\gamma|$ amplitude, averaged equally over all phase angles $\theta=\arg(\gamma)\in[-\pi,\pi]$. The right graphs, (d) to (f), is the fidelity for amplifying a coherent state with $\theta$ phase angle, with a fixed amplitude of $r=|\alpha|$. The solid orange lines were calculated using the chosen default parameter setting of $\{N=3,\alpha=0.5,g=2\}$, which can be compared to the dashed lines where one parameter of this set is changed. We can see in (a) and (d) that as the number of components $N$ increases, the fidelity increases because more phase angles can be amplified well (i.e. it becomes more linear due to overlapping components). From (b) and (e), we can see as the expected amplitude $\alpha$ becomes smaller, the protocol becomes more linear. Finally, (c) and (f) shows that as we demand more amplification gain $g$, we require more specific knowledge about the input state $\gamma$ to amplify with good fidelity.}
    \end{center}
\end{figure*}

\begin{figure*}[htb]
    \begin{center}
        \includegraphics[width=\linewidth]{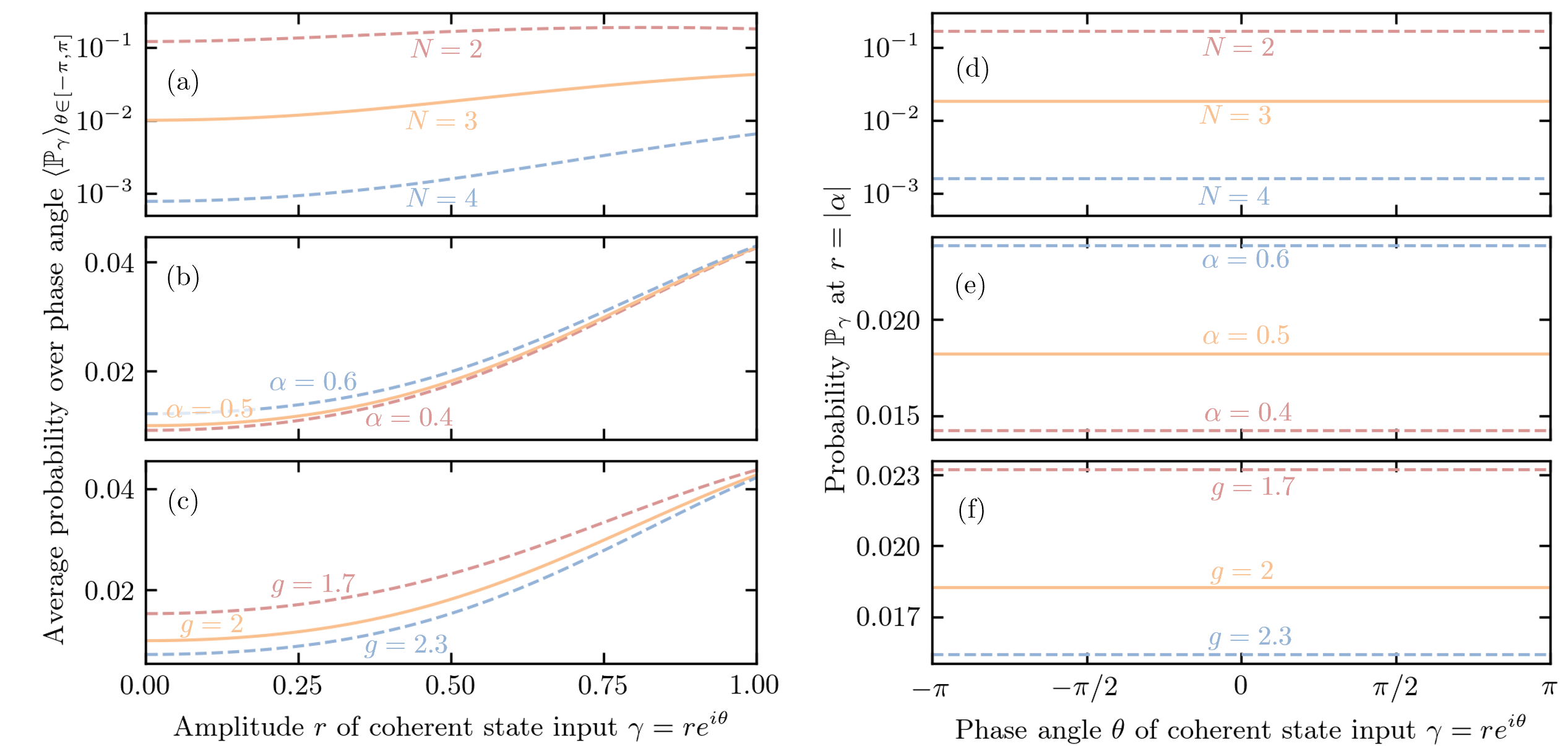}
        \caption{\label{fig:coherent-prob} 
            This is the same settings as Fig.~\ref{fig:coherent-fid}, except we are considering the success probability $\mathbb{P}_\gamma$ for amplifying $|\gamma\rangle$. Based on the right graphs, (d) to (f), it is clear that the input phase $\theta$ doesn't affect how likely we will get correct detection events. We can see that the parameter $N$ has the largest affect on the success probability, while $\alpha$ and $g$ has limited impact.}
    \end{center}
\end{figure*}

Here we consider our proposed amplifier given the input is just any arbitrary coherent state (i.e. it is not in one of the fixed places in phase space) as
\begin{align}
    |\gamma\rangle,\quad \gamma = q + ip = re^{i\theta},
\end{align}
assuming ideal conditions. 

Recall from Eq.~\eqref{eq:bobstate} that Bob prepares   
\begin{align}
    B_2(\tau) |0\rangle |\phi_{N,\beta}\rangle 
    &= \frac{1}{\sqrt{\mathcal{N}}} \sum_{b=1}^N \omega_N^b |-\omega_N^b\alpha\rangle |\omega_N^bg\alpha\rangle.
\end{align}
Note here $\alpha$ is now the \textit{expected} input amplitude that we set by choosing the cat resource amplitude $\beta$. Bob then sends $|-\omega_N^b\alpha\rangle$ towards Alice, who mixes this state on the $N$-splitter $S_N$ with $|\gamma\rangle$ resulting in
\begin{align}
    &\sum_{b=1}^N \frac{\omega_N^b}{\sqrt{\mathcal{N}}}  (S_N |\gamma\rangle |-\omega_N^b\alpha\rangle \otimes^N_{m=3} |0\rangle) |\omega_N^bg\alpha\rangle \nonumber \\ 
    &= \sum_{b=1}^N \frac{\omega_N^b}{\sqrt{\mathcal{N}}}  \otimes^N_{m=1} |(\gamma-\omega_N^{b+m-1}\alpha)/\sqrt{N}\rangle |\omega_N^bg\alpha\rangle. \label{eq:app_mixedcoh}
\end{align}
Alice will herald on the single-photon measurements $\langle0|\otimes^N_{m=2}\langle1|$, which requires the amplitude
\begin{align}
    &\langle0|\otimes^N_{m=2}\langle1|\ \otimes^N_{m=1} |(\gamma-\omega_N^{b+m-1}\alpha)/\sqrt{N}\rangle \nonumber \\ 
    &= e^{-\sum_{m=1}^N |\gamma-\omega_N^{b+m-1}\alpha|^2/(2N)} \prod_{m=2}^N \frac{\gamma-\omega_N^{b+m-1}\alpha}{\sqrt{N}} \nonumber \\ 
    &= \frac{e^{-(|\gamma|^2+|\alpha|^2)/2}}{N^{(N-1)/2}} \prod_{m=2}^N (\gamma-\omega_N^{b+m-1}\alpha) \nonumber \\ 
    &= \frac{e^{-(|\gamma|^2+|\alpha|^2)/2}}{N^{(N-1)/2}} d_b. \label{eq:app_ampcoh}
\end{align}
We simplified the exponent as follows
\begin{align}
    &\sum_{m=1}^N | \gamma - \omega^{b+m-1}_N \alpha |^2 \nonumber \\ 
    &= \sum_{m=1}^N (|\gamma|^2 + |\alpha|^2  - \omega^{b+m-1}_N \gamma^* \alpha -  \omega^{-b-m+1}_N \gamma \alpha^* ) \nonumber \\ 
    &= N(|\gamma|^2 + |\alpha|^2), 
\end{align}
where we used $\sum_{m=1}^N \omega_N^{m-1}= (1-\omega_N^N)/(1-\omega_N) = 0$. We also simplified the product as follows
\begin{align}
    d_b &= \prod_{m=2}^N (\gamma-\omega_N^{b+m-1}\alpha) \nonumber \\ 
    &= \prod_{m=1}^{N-1} (\gamma-\omega_N^{b+m}\alpha) \nonumber \\ 
    &= \omega_N^{b(N-1)}\alpha^{N-1}\prod_{m=1}^{N-1} \left(\frac{\gamma}{\omega_N^b\alpha}-\omega_N^m \right) \nonumber \\ 
    &= \omega_N^{b(N-1)}\alpha^{N-1}\sum_{m=1}^N \left(\frac{\gamma}{\omega_N^b\alpha}\right)^{m-1} \nonumber \\ 
    &= \omega_N^{b(N-1)}\alpha^{N-1} \frac{1-\gamma^N/(\omega_N^b\alpha)^N}{1 - \gamma/(\omega_N^b\alpha)} \nonumber \\ 
    &=  \frac{(\omega_N^b\alpha)^N-\gamma^N}{\omega_N^b\alpha - \gamma}, 
\end{align}
where we used Eq.~\eqref{eq:app_prod_3} in the fourth equality, and assumed that $\gamma \neq \omega_N^b\alpha$ to resolve the geometric sum in the fifth equality. If $\gamma =\omega_N^b\alpha$, then $d_b = \omega_N^{b(N-1)}\alpha^{N-1} N$ which when substituted in Eq.~\eqref{eq:app_ampcoh} we get an expression which is consistent with our previous Eq.~\eqref{eq:app_ampproof} result. 

By using the derived amplitude in Eq.~\eqref{eq:app_ampcoh}, we can see that applying these single-photon measurements to Eq.~\eqref{eq:app_mixedcoh} produces the unnormalised output state
\begin{align}
    |\psi'_{N,g\alpha}\rangle &= \frac{e^{-(|\gamma|^2+|\alpha|^2)/2}}{N^{(N-1)/2}\sqrt{\mathcal{N}}} \sum_{b=1}^N  \omega_N^b d_b |\omega_N^bg\alpha\rangle. 
\end{align}
We may then calculate the success probability of this protocol as
\begin{align}
    \mathbb{P}_\lambda &= \langle \psi'_{N,g\alpha} | \psi'_{N,g\alpha} \rangle \nonumber \\ 
    &= \frac{e^{-|\gamma|^2-|\alpha|^2}}{N^{(N-1)}\mathcal{N}} \sum_{a,b=1}^N \omega^{b-a}_N d_b d_a^* \langle\omega_N^ag\alpha|\omega_N^bg\alpha\rangle \nonumber \\ 
    &= \frac{e^{-|\gamma|^2-|\alpha|^2-|g\alpha|^2}}{N^{(N-1)}\mathcal{N}} \sum_{a,b=1}^N \omega^{b-a}_N d_b d_a^* e^{\omega^{b-a}_N|g\alpha|^2}. 
\end{align}
We can then calculate the fidelity of this protocol as
\begin{align}
    F_\lambda &= \frac{1}{\mathbb{P}_\lambda} |\langle g\gamma | \psi'_{N,g\alpha} \rangle|^2 \nonumber \\ 
    &= \frac{1}{\mathbb{P}_\lambda} \left|  \frac{e^{-(|\gamma|^2+|\alpha|^2)/2}}{N^{(N-1)/2}\sqrt{\mathcal{N}}} \sum_{b=1}^N  \omega_N^b d_b \langle g\gamma |\omega_N^bg\alpha\rangle \right|^2 \nonumber \\ 
    &= \frac{1}{\mathbb{P}_\lambda} \frac{e^{-|\gamma|^2-|\alpha|^2-|g\gamma|^2-|g\alpha|^2}}{N^{(N-1)}\mathcal{N}} \left| \sum_{b=1}^N  \omega_N^b d_b e^{\omega_N^bg^2\gamma^*\alpha} \right|^2 \nonumber \\ 
    &= \frac{e^{-|g\gamma|^2}\left| \sum_{b=1}^N  \omega_N^b d_b e^{\omega_N^bg^2\gamma^*\alpha} \right|^2 }{\sum_{a,b=1}^N \omega^{b-a}_N d_b d_a^* e^{\omega^{b-a}_N|g\alpha|^2}}. 
\end{align}
We plot these formulas in Fig.~\ref{fig:protocol-coherent} and Fig.~\ref{fig:coherent-fid} for fidelity, and in Fig.~\ref{fig:coherent-prob} for success probability. This is done for various parameter settings of our protocol, including the amount of coherent state components $N$, the expected amplitude $\alpha$, and the amplitude gain $g$. 

\section{Success Probability Analysis} \label{sec:prob}

If we assume an arbitrary input state which can be written in the $N$-components cat basis as in Eq.~\eqref{eq:input}, the unnormalised output state from our $N$-CT device is given by Eq.~\eqref{eq:output} as
\begin{align}
    |\psi_{N,g\alpha}\rangle = \frac{e^{-|\alpha|^2} \alpha^{N-1}}{N^{(N-3)/2}\sqrt{\mathcal{N}}} \sum_{a=1}^N c_a |\omega_N^a g\alpha\rangle.
\end{align}
Hence one can calculate the success probability as 
\begin{align}
    \mathbb{P} &= \langle \psi_{N,g\alpha}|\psi_{N,g\alpha} \rangle \nonumber \\
    &=  \frac{e^{-2|\alpha|^2} |\alpha|^{2(N-1)}}{N^{(N-3)}\mathcal{N}} \sum_{j=1}^N \sum_{k=1}^N c_j^* c_k e^{(\omega_N^{k-j} -1)|\alpha|^2 g^2}.
\end{align}
Since two coherent states are not orthogonal $\langle\beta|\alpha\rangle = e^{-(|\beta|^2+|\alpha|^2-2\beta^*\alpha)/2}$, the normalisation factor from the cat resource state $|\phi_{N,\beta}\rangle$ can be calculated as
\begin{align}
    \mathcal{N} &= \sum_{j,k=1}^N \omega_N^{k-j} \langle\omega_N^j\beta|\omega_N^k\beta\rangle \nonumber \\ 
    &= \sum_{j,k=1}^N \omega_N^{k-j} e^{(\omega_N^{k-j}-1)|\beta|^2} \nonumber \\
    &= \sum_{j,k=1}^N \omega_N^{k-j} e^{(\omega_N^{k-j}-1)|\alpha|^2(1/\eta+g^2)}.
\end{align}
Thus for an $N$-components input with known coefficients $c_a$ and amplitude $\alpha$, we can determine the probability for a given gain $g$ and channel transmissivity $\eta$.

Now, to gain an idea of how this probability scales, let us consider the coherent state input case $c_a=\delta_{a,a'}$. The success probability in this case is simply
\begin{align}
    \mathbb{P}_\text{c} &= \frac{e^{-2|\alpha|^2} |\alpha|^{2(N-1)}}{N^{(N-3)}\mathcal{N}}.
\end{align}
We have plotted this equation in Fig.~\ref{fig:protocol-loss}. Notice that $g^2$ and $1/\eta$ act similarly through the factor $\mathcal{N}$. If we increase $g$ or decrease $\eta$, then this requires increasing the resource $\beta=\alpha\sqrt{1/\eta+g^2}$. In the large amplitude limit $\beta\to\infty$ the cat resource $|\phi_{N,\beta}\rangle$ is the sum of $N$ orthogonal states, hence the normalisation factor simply becomes $\mathcal{N}\to N$. In other words,
\begin{align}
    \mathbb{P}_\text{lim,c} = \lim_{g\to\infty} \mathbb{P}_\text{c} = \lim_{\eta\to0} \mathbb{P}_\text{c} &= \frac{e^{-2|\alpha|^2} |\alpha|^{2(N-1)}}{N^{(N-2)}}.
\end{align}
By taking the derivative, we can calculate that the following input size
\begin{align}
    \alpha_\text{max} = \text{argmax}_\alpha \mathbb{P}_\text{lim,c} &= \sqrt{(N-1)/2},
\end{align}
maximises this success probability as
\begin{align}
    \mathbb{P}_\text{max,lim,c} = \max \mathbb{P}_\text{lim,c} \nonumber &= \frac{(2e)^{-(N-1)} (N-1)^{N-1}}{N^{(N-2)}} \nonumber \\ 
    &= \frac{2eN^2}{N-1} \left( \frac{N-1}{2eN} \right)^N.
\end{align}
Note we may include an additional $N$ factor due to Alice's multiple measurements as explained in Appendix~\ref{sec:measureproof}. Hence we have shown our $N$-CT device can teleamplify states with significant success probability in the large gain and/or large loss asymptotic limit. For example, for $N\mathbb{P}_\text{max,lim,c}(N=2) \approx 0.37$, $N\mathbb{P}_\text{max,lim,c}(N=3) \approx 0.14$, and $N\mathbb{P}_\text{max,lim,c}(N=5) \approx 0.01$.

\bibliography{paper}

\end{document}